\documentclass[12pt]{article}
\usepackage{amsmath,amssymb}
\begin{document}

\begin{titlepage}

\setlength{\baselineskip}{18pt}

                            \begin{center}
                            \vspace*{2cm}
        \large\bf  Entropy and  curvature variations from effective potentials  \\

                            \vfill

              \small\sf NIKOS \  KALOGEROPOULOS\\

                            \vspace{0.2cm}

 \small\sf  Department of Science,\\
            BMCC-The City University of New York\\
            199 Chambers St., New York, NY 10007, USA\\
                            \end{center}

                            \vfill

                     \centerline{\normalsize\bf Abstract}
                            \vspace{0.3cm}

\normalsize\rm\noindent \setlength{\baselineskip}{18pt} By using
the Jacobi metric of the configuration space, and assuming
ergodicity, we calculate the Boltzmann entropy $S$ of a
finite-dimensional system  around a non-degenerate critical point
of its potential energy $V$. We compare $S$ with the entropy of a
quantum or thermal system with effective potential $V_{eff}$. We
examine conditions, up to first order in perturbation theory,
under which these entropies are equal.

                             \vfill

                             \vfill
\noindent\rule{8cm}{0.2mm}\\
\begin{tabular}{ll}
\small\rm E-mail: & \small\rm nkalogeropoulos@bmcc.cuny.edu\\
                  & \small\rm nkaloger@yahoo.com
\end{tabular}
\end{titlepage}


                                 \newpage

\setlength{\baselineskip}{18pt}\rm

The Boltzmann entropy $S$ is a thermodynamic potential which is
particularly useful in describing the macroscopic behavior of a
system of constant energy $E$. We calculate the entropy variations
due to quantum or thermal effects of a finite dimensional
Hamiltonian system$^{1,2}$, where we assume that these
fluctuations are encoded in the effective potential$^{3,4,5}$
$V_{eff}$ of the system. $S$ is expressed as the logarithm of the
volume of the configuration space $\mathcal{M}$ of the system
under study. Calculating such a volume is impractical for most
systems, so to simplify the problem we confine ourselves to
spheres around a critical point $P$ of the classical potential
$V_c$. The evolution of a physical system in equilibrium takes
place in a neighborhood of a local minimum of $V_c$ compatible
with the external conditions, hence the importance of computations
taking place around such a critical point.\\

Determining volumes on a  manifold does not require the
introduction of a metric. The existence of a top-dimensional
differential form is sufficient, as in the case of the Liouville
form on phase space. However, if the manifold possesses a metric,
especially a Levi-Civita one, considerable simplifications occur
due to the strong constraints imposed upon it. On the other hand,
a physical model depends to a great extent on the metric chosen
for analyzing a system. Many times there is no a priori
justification of the choice of a specific metric. For our analysis
we follow essentially Krylov's ideas$^6$ and use the Jacobi metric
{\bf g}, which has components in a coordinate basis
\begin{equation}
               g_{ij}=2[E-V(x)]\delta_{ij}
\end{equation}
where $E$ is the total energy of the system and $V(x)$ is its
potential. From a mechanical viewpoint, the choice of this metric
is convenient since it is directly related to the kinetic energy
of the system. Another advantage of the Jacobi metric is that it
is reasonably easy to manipulate, since it is diagonal and
explicitly conformally flat. Because of its relation to the total
energy of the system $E$, ${\bf g}$ is ideal for analyzing a
system described macroscopically through its entropy $S$. The only
disturbing fact about the choice of ${\bf g}$ is that, it is
identically zero at the ``turning points", namely at subspaces of
the configuration space at which the kinetic energy is zero.
However, if all calculations take place away from the ``turning
points", ${\bf g}$ can be used reliably to analyze the behavior of
the system. In the present paper, using ${\bf g}$ we find that the
condition for $S$ to remain invariant under a perturbation, is
that the perturbing potential should satisfy a ``massive" elliptic
differential equation, and comment on its solutions in two extreme
cases, at
which considerable simplifications occur. \\

To begin with, an explicit expression for the Ricci scalar $R$ of
${\bf g}$ is needed.  The Christoffel symbols are given for a
general metric, in this coordinate basis,  by$^7$,
\begin{equation}
\Gamma^i_{jk}=\frac{1}{2}g^{im}\left( \partial_jg_{mk}+
  \partial_kg_{jm}-\partial_mg_{jk} \right)
\end{equation}
\noindent For ${\bf g}$ they become
\begin{equation}
\Gamma^i_{jk}=-\frac{1}{2(E-V)}\left(\delta^i_k\partial_jV+
\delta^i_j\partial_kV-\delta^{im}\delta_{jk}\partial_mV\right)
\end{equation}
The Riemann curvature tensor for any Levi-Civita metric is given,
in this coordinate basis, by
\begin{equation}
R^i_{ \ jkm} = \partial_k\Gamma^i_{jm}-\partial_j\Gamma^i_{km}
     +\Gamma^l_{jm}\Gamma^i_{kl}-\Gamma^l_{km}\Gamma^i_{jl}
\end{equation}
\noindent which, for ${\bf g}$ gives
\begin{eqnarray}
\lefteqn{R^i_{jkm}} \ \ \ \ \ & = &\frac{1}{2(E-V)^2}
                 ( \delta^i_k\partial_jV\partial_mV -
                 \delta^i_j\partial_kV\partial_mV+
                 \delta^{il}\delta_{jm}\partial_kV\partial_lV -
                 \delta^{il}\delta_{km}\partial_jV\partial_lV )+ \nonumber \\
              & & \frac{1}{2(E-V)}
                 ( \delta^i_k\partial_j\partial_mV-
                 \delta^i_j\partial_k\partial_mV+
                 \delta^{il}\delta_{jm}\partial_k\partial_lV-
                 \delta^{il}\delta_{km}\partial_j\partial_lV )+ \nonumber \\
             & & \frac{1}{4(E-V)^2}(
                 \delta^l_m\delta^i_k\partial_jV\partial_lV-
                 \delta_{km}\delta^{il}\partial_jV\partial_lV+
                 \delta^i_j\partial_kV\partial_mV+ \nonumber \\
             & & \delta^l_j\delta^i_k\partial_mV\partial_jV -
                 \delta^{il}\delta^{jm}\partial_kV\partial_lV-
                 \delta^i_k\delta^{rl}\delta_{jm}\partial_lV\partial_rV+
                 \delta^{ir}\delta_{jm}\delta^k_l\partial_lV\partial_rV- \nonumber \\
             & & \delta^i_j\delta^l_m\partial_kV\partial_lV+
                 \delta^{il}\delta_{jm}\partial_kV\partial_lV-
                 \delta^i_k\partial_jV\partial_mV-
                 \delta^i_j\delta^l_k\partial_mV\partial_lV+ \nonumber \\
             & & \delta^{il}\delta_{km}\partial_jV\partial_lV+
                 \delta^i_j\delta^{rl}\delta_{km}\partial_lV\partial_rV-
                 \delta^{in}\delta^l_j\delta_{km}\partial_lV\partial_nV)
\end{eqnarray}

\noindent Contraction of $i$ and $k$ gives the Ricci tensor
components
\begin{equation}
R_{i\!j}=
\frac{1}{2(E\!-\!V)}[(n\!-\!2)\partial_i\partial_jV\!+\!\delta_{ij}\triangle^2V]+
\frac{1}{4(E\!-\!V)^2}[3(n\!-\!2)\partial_iV\partial_jV\!-\!(n\!-\!4)\delta_{ij}
\Vert\nabla V\Vert^2]
\end{equation}

\noindent In (6) $\triangle$, $\nabla$ and $\Vert\cdot\Vert$ are
the Euclidean Laplacian, gradient and norm respectively. Here $n$
is the dimension of the configuration space $\mathcal{M}$. We
raise one index by using the inverse of the Jacobi metric ${\bf
g^{\rm -1}}$, and upon one further contraction we finally get the
Ricci scalar
\begin{equation}
R=\frac{n\!-\!1}{E\!-\!V}\triangle V -
\frac{(n\!-\!1)(n\!-\!6)}{4(E\!-\!V)^2}\Vert\nabla V\Vert^2
\end{equation}

\noindent One immediate observation is that the Jacobi metric
${\bf g}$ is not an Einstein metric, since it is, obviously,
impossible to find a constant $\Lambda$ such that $R_{ij}=\Lambda
g_{ij}$ even at the critical points of $V$. This does not allow us
to take advantage of the rather simple
variational characterization$^8$ of the Einstein metrics.\\

\noindent Our computations take place around a critical point $P$
of the classical/non-thermal potential $V_c(x)$. For simplicity,
we also assume that $P$ is non-degenerate. If it is degenerate, we
can use equivariant extensions of these arguments, especially if
the degenerate critical points are connected by a transitive, free
group action. For $P$, equations (6) and (7) for the Ricci tensor
and the Ricci scalar, respectively, simplify to
\begin{equation}
R_{i\!j}=
\frac{1}{2(E\!-\!V)}[(n\!-\!2)\partial_i\partial_jV\!+\!\delta_{ij}\triangle^2V]
\end{equation}
and
\begin{equation}
R=\frac{n\!-\!1}{2(E\!-\!V)}\triangle V
\end{equation}

\noindent We also present, for completeness, the calculation of
the volume of the image of a ball of radius $r$ in the
configuration space$^9$. Consider a normal coordinate system whose
origin is at $P$. Let $O_p$ denote the origin of a corresponding
coordinate system in $T_p\mathcal{M}$. We restrict ourselves to
balls $B_n(O_p,r)$ of radius $r$ in $T_p\mathcal{M}$ for which the
exponential map  \ $\exp_p: B_n(O_p,r)\rightarrow B_n(P,r)$  \ is
a diffeomorphism. In other words, we restrict ourselves inside the
injectivity radius of $P$. However, we have already assumed  in
our treatment that we want to be away from the ``turning points"
at which the metric vanishes degenerately. So, for practical
purposes in our treatment, $r$ is less than the minimum of the
injectivity radius of $P$ and the distance to the closest
``turning point" from $P$. We parametrize the points of \ $x \in
B_n(O_p,r)\setminus \{O_p\}$ \ of $T_pM$ in a polar coordinate
system as
\begin{equation*}
(\Vert x\Vert, \frac{x}{\Vert x\Vert })\in (0,r)\times S_{n-1}
\end{equation*}
where $S_{n-1}$ is the unit sphere in $n$ dimensions. Then the
calculation reduces to determining the ``radial" volume element in
$\mathcal{M}^{10}$. Consider an orthonormal basis $ \{ e_1, e_2,
\ldots, e_{n-1}, X \}$, where $X$ is the unit vector, with respect
to the canonical metric in $(0,r)$, which essentially defines a
point on $S_{n-1}$. Let us consider the geodesic $c(t)$ on
$\mathcal{M}$ with tangent vector $X$ and the transversal Jacobi
fields $J_1(t), J_2(t), \ldots, J_{n-1}(t)$ that satisfy \
$J_i(0)=0, \ DJ_i(0)=e_i, \ i=1,\ldots, n-1$. \ In this notation
$D$ stands for the unique Levi-Civita connection on $T\mathcal{M}$
compatible with ${\bf g}$. The ``radial" volume $\mu_t(X)$ is
given by the absolute value of the Jacobian of  \ $J_i, \
i=1,\ldots n-1$ \ or since $\mathcal{M}$ possesses a Riemannian
metric by
\begin{equation}
d\mu_t(X)=\sqrt{\det g(J_i, J_k)} \ dt, \ \ i,k=1,\ldots ,n-1
\end{equation}
Moreover \ $J_i(t)=tD[\exp (tX)e_i]$  \ and since \ $\{ e_i \}$, \
$ i=1\ldots, n-1$ \ is an orthonormal basis, we find that
\begin{equation}
d\mu_t(X)=t^{n-1} \ \sqrt{\det g(\exp(tX), \exp(tX))}
\end{equation}
Obviously  \ $d\mu_t(X)=t^{n-1}dt$ \ for $t=0$, which expresses
the fact that the Riemannian manifold, in normal coordinates, is
locally Euclidean. The volume of $B_n(P,r)$ is given by
\begin{equation}
vol B_n(P,r)=\int_{S_{n-1}}dS_{n-1}\int_0^r d\mu_t(X)
\end{equation}
In normal coordinates, it is well-known$^7$ that around point
$P\in \mathcal{M}$ the metric can be expressed in terms of the
Riemann curvature tensor and its covariant derivatives as a power
series with lowest order terms given by
\begin{equation}
g_{ij}(x)=\delta_{ij}+\frac{1}{3}R_{ikjm}(P)x^kx^m + O(\Vert
x\Vert^3)
\end{equation}
We recall$^{11}$, that the sectional curvature \ $K$ \ along a
plane spanned by the linearly independent vectors $Y=Y^iu_i$ and
$Z=Z^iu_i$ is a function on the Grassmann manifold \
$G_{2,n}(\mathcal{M})$ \  given by
\begin{equation}
K(Y,Z)=\frac{R_{ijkm}Y^iZ^jY^kZ^m}{(g_{ik}g_{jm}-g_{ij}g_{km})Y^iZ^jY^kZ^m}
\end{equation}
Here \ $\{ u_i \}, \ i=1, \ldots, n$ \ is a basis of
$T_p\mathcal{M}$, not necessarily orthonormal. The Ricci curvature
in the direction of $Y=Y^iu_i$ is the average of the sectional
curvatures along the planes containing $Y$, namely
\begin{equation}
\rho(Y)=\frac{R_{ij}Y^iY^j}{g_{kl}{Y^kY^l}}
\end{equation}
Using (13), we find
\begin{equation}
\det g_{ij}(tX)=1-\frac{1}{3}\rho(X)t^2+ O(t^3)
\end{equation}
so the infinitesimal radial volume element $d\mu_t(X)$, which is
proportional to \ $\sqrt{\det g_{ij}}$,  becomes in this
approximation
\begin{equation}
d\mu_t(X)=t^{n-1}-\frac{1}{6}\rho (X)t^{n+1}+O(t^{n+2})
\end{equation}
Therefore,  the volume of the ball $B_n(P,r)$ of $\mathcal{M}$ is
\begin{equation}
vol B_n(P,r)=
   \int_{S_{n-1}}dS_{n-1}\int_0^r(t^{n-1}-\frac{1}{6}\rho(X)t^{n+1})dt
\end{equation}
This non-trivial  integral essentially calculates the ``average"
of the Ricci curvature over all directions. We expect it,
therefore, to be proportional to the Ricci scalar $R$ at $P$. To
formally justify this, we consider an orthonormal basis \ $\{
e_1',\ldots e'_{n-1}\}$ \ in which the Ricci curvature, considered
as a quadratic form, is diagonal. Then the elements \ $e'_i, \
i=1,\ldots n-1$ \ of this basis are eigenvectors of the Ricci
curvatures with eigenvalues \ $s_i, \ i=1,\ldots n-1$. \ In this
basis,
\begin{equation}
X=\sum_{i=1}^{n-1}X^ie'_i
\end{equation}
so the Ricci curvature and the Ricci scalar, respectively, are
expressed as
\begin{equation}
\rho(X)=\sum_{i=1}^{n-1}s_i(X^i)^2, \hspace{1cm}
R_p=\sum_{i=1}^{n-1} s_i
\end{equation}
The volume $\omega_n$ of the unit ball in Euclidean space is
given, in spherical coordinates, by
\begin{equation}
  \omega_n=\int_{\sum_{i=1}^{n-1}(X^i)^2=1}(X^i)^2dS_{n-1}
\end{equation}
which gives
\begin{equation}
 \int_{S_{n-1}}\rho(X)dS_{n-1}=R_p\omega_n
\end{equation}
Moreover,
\begin{equation}
 vol
 B_n(P,r)=\int_{S_{n-1}}dS_{n-1}\int_0^r\left[t^{n-1}-\frac{1}{6}\rho(X)t^{n+1}\right] \ dt
\end{equation}
which becomes
\begin{equation}
vol B_n(P,r)=\frac{1}{n}r^nA_{n-1}-\frac{r^{n+2}}{6(n+2)}R_p \
\omega_n
\end{equation}

\noindent It is also well-known$^{10}$ that  the volume of a ball
of radius $r$, is given for any  \ $n\in\mathbb{N}$ \ by
\begin{equation}
\omega_n=\frac{\pi^\frac{n}{2}}{\Gamma(\frac{n}{2}+1)}
\end{equation}
Since  \ $d\Omega_n=\alpha_{n-1}dr$ \ we find for the area of the
unit sphere $S_{n-1}$ that
\begin{equation}
\alpha_{n-1}=n\omega_n
\end{equation}
Putting (9), (24), (25), (26) together, we finally find for the
volume of the geodesic ball $B_n(P,r)$
\begin{equation}
 vol B_n(P,r)=\frac{\pi^{\frac{n}{2}}r^n}{\Gamma (\frac{n}{2}+1)}
              -\frac{\pi^{\frac{n}{2}}r^{n+2}}{6(n+2)\Gamma
              (\frac{n}{2}+1)}R_p
\end{equation}
The Boltzmann entropy is defined by
\begin{equation}
S=k_B\ln volB_n(P,r)
\end{equation}
where $k_B$ is Boltzmann's constant. This definition has physical
significance as long as the system is ergodic, i.e. as long as the
geodesic flow in  $\mathcal{M}$ of the system is ergodic. For such
a system, combining (27) and (28) we find
\begin{equation}
 S=k_B\ln \left\{ \frac{\pi^{\frac{n}{2}}r^n}{\Gamma (\frac{n}{2}+1)}
              -\frac{\pi^{\frac{n}{2}}r^{n+2}}{6(n+2)\Gamma
              (\frac{n}{2}+1)}R_p \right\}
\end{equation}

\noindent Let's assume that the renormalized effective potential
of the system is
\begin{equation}
V_{eff}(x)=V_c(x)+\lambda\tilde{V}(x)
\end{equation}
Here $\lambda$ indicates a small coupling constant which is used
as a  parameter in a power series expansion. \ $\tilde{V}(x)$ \
represents the quantum/thermal contribution to the classical
potential. The corresponding Jacobi metric on $\mathcal{M}$ is
\begin{equation}
 \tilde{g}_{ij}=2[E-V_c(x)-\lambda\tilde{V}(x)]\delta_{ij}
\end{equation}
 We observe that both metrics are conformally equivalent, and
$\bf \tilde{g}$ can be written as
\begin{equation}
 \tilde{g}_{ij}=2\left[ E-V_c(x)\right]\left[ 1-\lambda
 \frac{\tilde{V}(x)}{E-V_c(x)}\right]\delta_{ij}
\end{equation}
which means
\begin{equation}
 \tilde{g}_{ij}=\left[ 1-\lambda \frac{\tilde{V}(x)}{E-V_c(x)}\right] g_{ij}
\end{equation}
The conformal factor relating these metrics is therefore
\begin{equation}
 \sigma(x)=\frac{1}{2}\ln\left[ 1-\lambda\frac{\tilde{V}(x)}{E-V_c(x)}\right]
\end{equation}
A laborious computation, done either straightforwardly in
coordinates, or in a coordinate free way using Nomizu's
approach$^{13}$ , gives
\begin{equation}
\tilde{R}=e^{-2\sigma(x)}\left[ R-2(n-1)B_i^i\right]
\end{equation}
where
\begin{equation}
B_i^i=\left(\frac{n}{2}-1\right)g^{kl}\partial_k\sigma\partial_l\sigma
+
g^{kl}\left(\partial_k\partial_l\sigma-\Gamma_{kl}^m\partial_m\sigma\right)
\end{equation}
At the origin of the normal coordinate system, the Christoffel
symbols are zero. Therefore $B_i^i$ at $P$  becomes
\begin{eqnarray*}
B_i^i & = & \frac{1}{4(E-V_c-\lambda\tilde{V})}
[\lambda^2(\frac{n}{2}-1)g^{kl}\partial_k\tilde{V}\partial_l\tilde{V}-
2\lambda g^{kl}(E-V_c-\lambda\tilde{V})\partial_k\partial_l\tilde{V}\\
    &  &
-2\lambda\tilde{V}g^{kl}(1-\lambda\frac{\tilde{V}}{E-V_c})\partial_k\partial_lV_c
+2\lambda^2g^{kl}\partial_k\tilde{V}
\partial_l\tilde{V}]
\end{eqnarray*}
Expanding in powers of $\lambda$, and keeping up to linear terms
in this parameter, we get
\begin{equation}
B_i^i=-\frac{\lambda}{2(E-V_c)}\left[
   g^{kl}\partial_k\partial_l\tilde{V}+
      \frac{\tilde{V}}{E-V_c}g^{kl}\partial_k\partial_lV_c\right]
\end{equation}
This gives
\begin{equation}
R_p-\tilde{R}_p=-2(n-1)\frac{\lambda}{2(E-V_c)}\left[
g^{kl}\partial_k\partial_l\tilde{V}+
  \frac{\tilde{V}}{E-V_c}g^{kl}\partial_k\partial_lV_c\right]-
  \lambda\frac{\tilde{V}}{E-V_c}R_p
\end{equation}
We observe that to zeroth order in $\lambda$
\begin{equation}
 R_p=\tilde{R}_p
\end{equation}
so the difference between the entropies $S$ and $\tilde{S}$ of the
original and the perturbed system, respectively, due to (29)
becomes
\begin{equation}
\tilde{S}-S=k_B\ln\left[
1-\frac{r^2}{6(n+2)}\cdot\frac{R_p-\tilde{R}_p}{1-\frac{r^2}{6(n+2)}R_p}\right]
\end{equation}
Substituting (38) into (40), expanding the logarithm  in powers of
$\lambda$ and keeping up to linear terms,  we  find
\begin{eqnarray}
 \tilde{S}-S & = & k_B(-\lambda)\left[1-\frac{r^2}{6(n+2)}R_p\right]^{-1}
                          \frac{r^2}{6(n+2)} \nonumber \\
  &  &  \left\{ \frac{n-1}{E-V_c}\left[
g^{kl}\partial_k\partial_l\tilde{V}+
  \frac{\tilde{V}}{E-V_c}g^{kl}\partial_k\partial_lV_c\right] +
                \frac{\tilde{V}}{E-V_c}R_p \right\}
\end{eqnarray}

\noindent In this approximation, the entropy remains invariant if
\begin{equation}
 \frac{n-1}{E-V_c}\left[ g^{kl}\partial_k\partial_l\tilde{V}+
  \frac{\tilde{V}}{E-V_c}g^{kl}\partial_k\partial_lV_c\right] +
                \frac{\tilde{V}}{E-V_c}R_p = 0
\end{equation}
which reduces to
\begin{equation}
3\frac{\tilde{V}(x)}{E\!-\!V_c(x)}R_p+\frac{n\!-\!1}{E\!-\!V_c(x)}
\triangle\tilde{V}(x)=0
\end{equation}
which eventually gives
\begin{equation}
 \triangle\tilde{V}(x)+\frac{3R_p}{n-1}\tilde{V}(x)=0
\end{equation}
Let $\triangle_g$ denote the scalar Laplacian of the configuration
space with respect to the Jacobi metric $\bf g$. Then$^7$
\begin{equation}
 \triangle_g=\frac{1}{\sqrt{g}}\partial_i(g^{ij}\sqrt{g}\partial_j)
\end{equation}
At $P$, we find  straightforwardly, for $\tilde{V}(x)$
\begin{equation}
 \triangle\tilde{V}(P)= 2(E-V_c(x))\triangle_g\tilde{V}(P)
\end{equation}
From (9) we can also see that the Ricci scalar at $P$ can be
expressed as
\begin{equation}
 R_p=(n-1)\triangle_gV_c(P)
\end{equation}
 This condition can be equivalently written in
terms of the scalar Laplacian $\triangle_g$ associated with $\bf
g$, as
\begin{equation}
 \triangle_g\tilde{V}(x)+\frac{3\triangle_gV_c(P)}{2(E-V_c(P))}\tilde{V}(x)=0
\end{equation}
Therefore invariance of the Boltzmann entropy under perturbations
amounts to the perturbing potential $\tilde{V}(x)$ obeying a
``massive" Laplace equation with ``mass"
\begin{equation}
  M=\frac{3|\triangle_gV_c(P)|}{2(E-V_c(P))}
\end{equation}
If $P$ is a  relative minimum of $V_c(x)$, then $M$ is
positive-definite. Indeed, $P$ is a local minimum of $V_c(x)$ so
the Hessian $H_{ij}$ at $P$ is
\begin{equation}
  H_{ij}V_c(P)=\partial_i\partial_jV_c(P)
\end{equation}
and is positive definite. By definition $^{8,9,11}$
\begin{equation}
  tr H_{ij}=\triangle_g
\end{equation}
 so after diagonalizing $H_{ij}$, we see that $\triangle_gV_c(P)>0$.\\
A special case arises if
\begin{equation}
 E=\frac{6}{n-2}+V_c(P)
\end{equation}
Then the equation for $\tilde{V}(x)$ inside the geodesic ball
becomes
\begin{equation}
\left[\triangle_g+\frac{n-2}{4(n-1)}R_p\right]\tilde{V}(x)=0
\end{equation}
which means that $\tilde{V}(x)$ is an eigenfunction of the
conformal Laplacian$^8$
\begin{equation}
 \triangle_g+\frac{n-2}{4(n-1)}R_p
\end{equation}
corresponding to zero eigenvalue. In case $E$ has any other value,
$\tilde{V}(x)$ is a solution of
\begin{equation}
\left[\triangle_g+\frac{n-2}{4(n-1)}R_p\right]\tilde{V}(x)=
\left[\frac{n-2}{4(n-1)}R_p-\frac{3\triangle_gV_c}{2(E-V_c)}\right]\tilde{V}(x)
\end{equation}
This is very similar in form, though just a linearized equation,
to the Euler-Lagrange equation arising in the variational
determination of the Yamabe constant$^{15}$ of a conformal class
of metrics on a manifold $\mathcal{M}$. The common point of these
equations is that both arise from variations within a conformal
class of metrics of $\mathcal{M}$.\\


If \ $E\rightarrow\infty$, \ then the Riemann tensor (5) and
consequently the Ricci tensor (6) approach zero. In this limit
$\mathcal{M}$ is Euclidean  and the induced Riemannian measure on
it, i.e. the induced volume element, is uniform. Then the
effective mass $M$ in (48) is zero. This limit describes the
evolution of a system of very high kinetic energy. Because of the
high kinetic energy, as the system evolves in $\mathcal{M}$, it is
unaffected by any curvature deviations of $\mathcal{M}$ from
flatness. For $S$ to remain invariant under perturbations, to
first order in $\lambda$, the perturbing potential $\tilde{V}(x)$
must be a harmonic function with respect to the Jacobi metric.
Such a function depends on the boundary conditions imposed on the
sphere \ $\partial B_n(P,r)$. \ If  $r$ is small enough, it is
reasonable to assume that the value of $V(x)$ at all points $x$ of
\ $\partial B_n(P,r)$ \ is the same, i.e. that
\begin{equation}
\tilde{V}(x)|_{\partial B_n(P,r)}=V_o
\end{equation}
The unique harmonic function satisfying this condition is
the constant function $V_o$.\\

\noindent The other limiting case is  one in which \ $E\rightarrow
V_c(P)$. \ Then the Riemann tensor (5) as well as the Ricci tensor
(6) approach infinity, a fact that indicates the curvature tensors
develop conical or cuspidal singularities, depending on the exact
form of $V_c(P)$. The source of the singular behavior is that in
this limit the Jacobi metric becomes exactly zero. Therefore it is
not a reliable probe of the structure of $\mathcal{M}$ and the
description of the evolution of the system. In such a case, the
system is evolving on $\mathcal{M}$ around $P$ having a  very
small kinetic energy. No curvature fluctuations of $\mathcal{M}$
can be overcome by the system, exactly because of  its small
kinetic energy. Equivalently, we observe from (49) that $M$
becomes very high. For the energy $E$ to remain conserved the
system has to confine itself in a very small ball $B_n(P,r)$ of
$\mathcal{M}$. In this limit, the Dirichlet problem implies that
the fluctuating potential should vanish at $\partial B_n(P,r)$.
The unique solution to the boundary-value problem is therefore
\begin{equation}
 \tilde{V}(x)=0
\end{equation}
so $S$ remains invariant as long as the first order radiative
corrections of $V_{eff}$ are trivial.\\

\noindent These two limiting cases reflect the fact that the only
harmonic functions on a compact manifold are constants. For all
other values of $E$, between zero and infinity the system
interpolates between these two extreme cases. Depending on the
value of $r$, $R_p$ and the sign of $R_p$, the accessible part
$\mathcal{M}$ may be disconnected, or may even be topologically
non-trivial. This fact places further
restrictions on the form of $\tilde{V}(x)$.\\

When  \ $R_p>0$ \ there is no guarantee that the system is ergodic
even on $B_n(P,r)$ for small $r$. Ergodicity of the geodesic flow
associated to ${\bf g}$ is related to the Riemann curvature tensor
(5) or equivalently to the sectional curvature $K$ (14) of ${\bf
g}$ rather than its Ricci scalar $R_p$ (9). There are various
quantitative measures of the complexity of the geodesic flow of a
manifold. One of the most studied measures is the topological
entropy ${h_{top}}^{16}$. It is known$^{16}$ that if the sectional
curvature $K$ (14) of a manifold $\mathcal{M}$ is everywhere
negative then its fundamental group has exponential growth$^{17}$,
the geodesic flow is ergodic and the topological entropy is
positive. $K$ is not negative, generically, for ${\bf g}$ on
$\mathcal{M}$ as can be seen from (14). Confining our attention to
the sphere under consideration, it is impossible to have negative
$K$ everywhere inside $B_n(P,r)$ for any value of $r$. This is a
direct consequence of Milnor's theorem$^{17}$ and the fact that \
$\pi_1(B_n(P,r))$ \ is trivial for \ $n>1$. A metric of everywhere
positive $K$ however may or may not have $h_{top}>0$. Since $K$
does not have a constant sign throughout $B_n(P,r)$, for the
entropy $S$ to be a physically relevant measure of disorder,
we have to assume ergodicity of the geodesic flow of ${\bf g}$.\\

A by-product of these calculations is a contribution towards the
computation of the Euler characteristic needed in the topological
hypothesis$^6$. According to it, the source of phase transitions
lies in the topology change of appropriate sub-manifolds
$\mathcal{N}$ of $\mathcal{M}$. This topology change is detected
by calculating the Euler characteristic of $\mathcal{N}$ usually
by using Morse theory$^6$. Let $\overline{\mathcal{N}}$ be a
$d-$fold un-ramified covering of $\mathcal{N}$. Then$^{18}$
$\chi(\overline{\mathcal{N}})= d\chi(\mathcal{N})$, so the
calculation of the Euler characteristic of $\mathcal{N}$ reduces
to that of $\overline{\mathcal{N}}$ where the integration measure
in the Gauss-Bonnet-Chern formula$^{19}$ applied to
$\overline{\mathcal{N}}$ is provided by $(27)$ as long as
$\overline{\mathcal{N}}$ has the
topological type of a homology $m-$sphere, \ $m\leq n$.\\

                   \vspace{1cm}


\centerline{\sc References}

\begin{enumerate}
 \item V.I. Arnold, Mathematical Methods of Classical Mechanics,
       2nd Ed., Springer-Verlag 1997.
 \item J.V. Jose, E. Saletan,  Classical Dynamics, Cambridge
       Univ. Press 1998.
 \item V.N. Popov, Functional Integrals and Collective
       Excitations, Cambridge Univ. Press 1990
 \item S. Pokorski, Gauge Field Theories, Cambridge Univ. Press 1989
 \item S. Weinberg, The Quantum Theory of Fields, Vols. I, II,
       Cambridge Univ. Press 1996
 \item L. Casetti, M. Pettini, E.G.D. Cohen, Geometric Approach to
       Hamiltonian Dynamics and Statistical Mechanics, Phys. Reports 337, 237 (2000)
 \item S. Kobayashi, K. Nomizu, Foundations of Differential
       Geometry, Vols. I, II, \ John Wiley \& Sons 1963
 \item A. Besse, Einstein Manifolds, Springer-Verlag 1987
 \item S. Gallot, D. Hulin, J. Lafontaine, Riemannian Geometry,
       Springer-Verlag 1987
 \item H. Federer, Geometric Measure Theory, Springer-Verlag 1969
 \item J. Jost, Riemannian Geometry and Geometric Analysis, 2nd
       Ed., Springer-Verlag 1998
 \item J. Jost, Postmodern Analysis, Springer-Verlag 1998
 \item E.T. Whittaker, G.N. Watson, A Course in Modern Analysis,
       4th Ed., Cambridge Univ. Press 1927
 \item M. Nakahara, Geometry, Topology and Physics, IOP Publishing 1990
 \item T. Aubin, Some Nonlinear Problems in Riemannian Geometry,
       Springer-Verlag 1998
 \item A. Katok, B. Hasselblatt, Introduction to the Modern Theory
       of Dynamical Systems, Cambridge Univ. Press 1997
 \item J. Milnor, A note on curvature and the fundamental group,
       J. Diff. Geom. 2, 1-7, 1978
 \item M.J. Greenberg, J.H. Harper, Algebraic Topology, A First
       Course, Addison-Wesley Publishing Co. 1981
 \item P. Gilkey, Invariance theory, the heat equation and the Atiyah-Singer
       index theorem, Publish or Perish Inc. 1984
\end{enumerate}

                               \vfill

\end{document}